\documentclass[]{aa}
\usepackage{times,epsfig} 

\newcommand{\U}{\mbox{$\mathit{U}$}}
\newcommand{\B}{\mbox{$\mathit{B}$}}
\newcommand{\V}{\mbox{$\mathit{V}$}}
\newcommand{\R}{\mbox{$\mathit{R}$}}
\newcommand{\I}{\mbox{$\mathit{I}$}}
\newcommand{\J}{\mbox{$\mathit{J}$}}
\newcommand{\HH}{\mbox{$\mathit{H}$}}
\newcommand{\K}{\mbox{$\mathit{K}$}}
\newcommand{\UBVRI}{\mbox{$\mathit{UBVRI}$}}
\newcommand{\BVRI}{\mbox{$\mathit{BVRI}$}}

\newcommand{\BRI}{\mbox{$\mathit{BRI}$}}

\newcommand{\JHK}{\mbox{$\mathit{JHK}$}}
\newcommand{\EBV}{\mbox{$\mathit{E(B-V)}$}}
\newcommand{\BmV}{\mbox{$\mathit{B-V}$}}

\begin{document}
\newcommand{\dm}{{\ensuremath{\Delta m_{15}}}}

\thesaurus{03(08.19.4; 08.06.3)}
\title{Epochs of Maximum Light and Bolometric Light Curves of Type Ia
Supernovae} 
\titlerunning{Bolometric Light Curves of Type Ia Supernovae}
\author{G.~Contardo\inst{1,2} \and B.~Leibundgut\inst{1} \and
W.~D.~Vacca\inst{3}}   
\mail{contardo@mpa-garching.mpg.de}
\institute{European Southern Observatory, Karl-Schwarzschild-Strasse 2,
D-85748 Garching, Germany \and Max--Planck--Institut f\"ur
Astrophysik, Karl-Schwarzschild-Strasse 1, D-85748 Garching, Germany
\and Institute for Astronomy, University of Hawaii, 2680 Woodlawn Drive,
Honolulu, HI 96822, USA} 
\date{Received 24 March 2000; accepted 12 May 2000}
\maketitle

\begin{abstract} 
We present empirical fits to the {\UBVRI} light curves of type Ia supernovae.
These fits are used to objectively evaluate light curve parameters.  We find
that the relative times of maximum light in the filter passbands are very
similar for most objects. Surprisingly the maximum at longer wavelengths is
reached earlier than in the {\B} and {\V} light curves.  This clearly
demonstrates the complicated nature of the supernova emission.

Bolometric light curves for a small sample of well-observed SNe~Ia are
constructed by integration over the optical filters. In most objects a plateau
or inflection is observed in the light curve about 20$-$40 days after
bolometric maximum.  The strength of this plateau varies considerably among the
individual objects in the sample. Furthermore the rise times show a range of
several days for the few objects which have observations early enough for such
an analysis. On the other hand, the decline rate between 50 and 80 days past
maximum is remarkably similar for all objects, with the notable exception of
SN~1991bg.  The similar late decline rates for the supernovae indicate that the
energy release at late times is very uniform; the differences at early times
is likely due to the radiation diffusing out of the ejecta.

With the exception of SN~1991bg, the range of absolute bolometric luminosities
of SNe~Ia is found to be at least a factor of 2.5. The nickel masses derived
from this estimate range from 0.4 to 1.1 $M_{\sun}$. It seems impossible to
explain such a mass range by a single explosion mechanism, especially since the
rate of $\gamma-$ray escape at late phases seems to be very uniform.

\end{abstract} 

\section{Introduction}

The temporal evolution of a supernova's luminosity contains important
information on the physical processes driving the explosion.  The peak
luminosity of a Type Ia Supernova (SN~Ia) is directly linked to the amount of
radioactive \element[][56]{Ni} produced in the explosion (Arnett et al.\ \cite{arn85},
Branch \& Tammann \cite{bra92}, H\"oflich et al.\ \cite{hoe97}, Eastman
\cite{eas97}, Pinto \& Eastman \cite{pin00a}). The rise time of the light curve
is determined primarily by the explosion energy and the manner in which the
ejecta become optically thin to thermalized radiation, i.e. the opacity
(Khokhlov et al.\ \cite{kho93}). The late decline of the light curve is
governed by the combination of the energy input by the radioactive material and
the rate at which this input energy is converted to optical photons in the
ejecta (Leibundgut \& Pinto \cite{lei92}).

The apparent uniformity of SN~Ia light curves in photographic ({\it pg}), {\B},
and {\V} filters (Minkowski \cite{min64}) prompted the adoption of standard
light curve templates (e.g. Elias et al.\ \cite{eli85}, Doggett \& Branch
\cite{dog85}, Leibundgut et al.\ \cite{lei91b}, Schlegel \cite{schle95}).
Early indications that the standard templates fail to describe the full range
of SN~Ia light curves came from the observations of SN~1986G, which displayed a
much more rapid evolution than any other SN observed up to that time (Phillips
et al.\ \cite{phi87}). The demise of the simple standard candle treatment was
brought about by the observations of the faint SN~1991bg (Filippenko et al.\
\cite{fil92a}, Leibundgut et al.\ \cite{lei93}, Turatto et al.\ \cite{tur96a})
and the subsequent derivation of a correlation between peak luminosity and
decline rate after maximum (Phillips \cite{phi93}, Hamuy et al.\ \cite{ham96d},
Riess et al.\ \cite{rie96b}).  A clear demonstration that SNe~Ia do not display
a uniform photometric evolution was provided earlier from the infrared {\J},
{\HH}, and {\K} light curves (Elias et al.\ \cite{eli85}, Frogel et al.\
\cite{fro87}).  Observations and analyses of the near-IR {\R} and {\I} light
curves (Suntzeff \cite{sun96}, Vacca \& Leibundgut \cite{vac97}) confirm this
variation.  The red and near-infrared light curves exhibit a second maximum
$\sim$ 20 to 30 days after the {\B} peak. This second maximum occurs at
different phases and with differing strengths in individual SNe~Ia and in at
least one case (SN~1991bg; Filippenko et al.\ \cite{fil92a}, Leibundgut et al.\
\cite{lei93}, Turatto et al.\ \cite{tur96a}) is altogether absent.

The light curves of SNe~Ia are often described as a one-parameter family. The
correlation of the {\BVRI} light curve shapes
with the peak absolute magnitudes has been employed to improve the distance
measurements derived from SNe~Ia (Hamuy et al.\ \cite{ham96d}, Riess et al.\
\cite{rie96b}, Garnavich et al.\ \cite{gar98}, Schmidt et al.\ \cite{sch98},
Riess et al.\ \cite{rie98a}) and is of fundamental importance to keep the
systematic uncertainties in the derivation of cosmological parameters small.
Techniques that fit standard templates (e.g., Hamuy et al.\ \cite{ham96d}), or
modified versions of these templates (e.g., Riess et al.\ \cite{rie96b};
Perlmutter et al.\ \cite{per97}), to the observed light curves make use of this
one parameter description of the light curves.  These methods have the
advantage that they can be applied even to rather sparsely sampled light
curves.  In a more extreme form, the photometry can be supplemented by
spectroscopy to provide a distance measurement with minimal data coverage. This
``snapshot'' method has been advocated by Riess et al.\ (\cite{rie98b}). All
these methods make the assumption that SNe~Ia form an ``ordered class''.  This
description is validated by the improvement in the scatter around the linear
expansion line in the local Universe and also by the fact that new objects can
be successfully corrected with the correlations derived from an independent
sample.

While clearly useful for comparing local SNe with high redshift SNe, the
template method does not allow one to investigate finer and more individual
features in a large sample of SN~Ia light curves. Hence, the detailed study of
the explosion and radiation physics cannot be carried out with such an
analysis.  For data sets which are densely sampled, however, template methods
are not necessary. Recent bright SNe~Ia have been observed extensively and very
detailed, and accurate light curves have become available.  Most of these
supernovae have been used as the defining objects for the templates to correct
other, more sparsely observed, SNe~Ia.

To analyze light curves of many SNe~Ia in an individual fashion a parameter
fitting method which can be applied to single filter light curves has been
devised (Vacca \& Leibundgut \cite{vac96,vac97}).  The photometric data are
approximated by a smooth fitting function. We are not fitting model light
curves based on explosion physics, but simply attempt to match the data in an
objective way. We have investigated well-observed SNe in a small sample to
check our method. It allows us to search for correlations among various light
curve parameters, to accurately fit the filter light curves at any phase, and
to construct continuous bolometric light curves. The application to the light
curves of SN~1994D has already provided one of the first bolometric light
curves of SNe~Ia (Vacca \& Leibundgut \cite{vac96}).

Observationally-derived bolometric light curves provide a measure of the total
output of converted radiation of SNe~Ia, and therefore serve as a crucial link
to theoretical models and calculations of SN explosions and evolution. The
total luminosity from a SN~Ia is much easier to calculate from theoretical
models than the individual filter light curves, which are dominated by line
blending effects requiring complicated multi-group calculations (Leibundgut \&
Pinto \cite{lei92}, Eastman \cite{eas97}, H\"oflich et al.\ \cite{hoe97}). In
addition, the observed bolometric peak luminosity of SNe~Ia provides a measure
of the total amount of nickel synthesized in the explosion and can be used to
test various explosion models.  Although no SN~Ia has ever been observed in
every region of the electromagnetic spectrum simultaneously, fortunately
bolometric light curves can be constructed almost entirely from optical data
alone (Suntzeff \cite{sun96}, Leibundgut \cite{lei96a}).

In Sect.~\ref{data} we describe the sample of objects and the data set used.
The parameter fitting method employed for our analysis is described in
Sect.~\ref{fitlc}.  This is followed by a discussion of the phases of the
maximum epoch in different filters (Sect.~\ref{epmax}). The construction of the
bolometric light curves is presented in Sect.~\ref{constrbol} followed by a
discussion of the uncertainties in the bolometric light curves.  We summarize
our findings in Sect.~\ref{disc}; our conclusions are given in
Sect.~\ref{conc}.

\section{Observational Data \label{data}}

The data for our analysis come from recent photometric observations of bright
SNe~Ia, consisting of the large data collections presented by Hamuy et al.\
(\cite{ham96b}) and Riess (\cite{rie96c}, see also Riess et al.\ \cite{rie99}).
Table \ref{data.tab} summarizes the objects which have sufficient data, in at
least the \BVRI\ filters, to construct accurate individual filter light curves.
Listed are the available filters (column 2) and the references of the SN
photometry (column 3), the adopted distance modulus (column 4) and its source
(column 5). The distance modulus is taken either from direct distance
measurements via cepheids, surface brightness fluctuations or Tully-Fisher
luminosity--line width relation or by adopting a $H_0$ of
65~km~s$^{-1}$~Mpc$^{-1}$ for galaxies in the Hubble flow
($v>2500$~km~s$^{-1}$).  We use the host galaxy dust extinction values of
Phillips et al.\ (\cite{phi99}) (column 6). The Galactic extinction given by
the COBE dust maps is provided in column 7 (Schlegel et al.\ \cite{schle98}).

Not all objects have sufficient observations in all filters to guarantee that
our method will work. In some cases, the light curve fits had to be restricted
to a limited phase range.  Only three objects (SN~1989B, SN~1991T, and
SN~1994D) have a significant number of {\U} filter observations. Our analysis
does not include any {\JHK} or ultraviolet data.

%
%
\begin{table*}
\caption{Photometry of well-observed SN~Ia \label{data.tab}} 
\begin{minipage}{\columnwidth}
\begin{tabular}{l  r l  c r  c c} 
SN & \multicolumn{1}{c}{Filter} & ref.$^a$
 & DM & ref.$^a$\  & \EBV$_\mathrm{host}$  & 
\multicolumn{1}{c}{\EBV$_\mathrm{gal}$} \\ 
    &     &     &     &     &  \multicolumn{1}{l}{Phillips et al.\
 (\cite{phi99})}   &    
\multicolumn{1}{l}{Schlegel et al.\ (\cite{sch98})} \\
(1) & \multicolumn{1}{c}{(2)} & (3) & (4) & 
\multicolumn{1}{c}{(5)} & (6) & 
(8)  \\                                        
\hline 					       
SN1989B  & {\UBVRI} & 1            & 30.22   & 14~ &    0.340 &     0.032 \\
SN1991T  & {\UBVRI} & 2            & 31.07   & 15~ &    0.140 &     0.022 \\
SN1991bg & {\BVRI}  & 3, 4, 5      & 31.26   & 16~ &    0.030 &     0.040 \\
SN1992A  & {\BVRI}  & 6            & 31.34   & 17~ &    0.000 &     0.017 \\
SN1992bc & {\BVRI}  & 7            & 34.82   &  7~ &    0.000 &     0.022 \\
SN1992bo & {\BVRI}  & 7            & 34.63   &  7~ &    0.000 &     0.027 \\
SN1994D  & {\UBVRI} & 8, 9, 10, 11 & 30.68   &  9~ &    0.000 &     0.022 \\
SN1994ae & {\BVRI}  & 12           & 31.86   & 18~ &    0.120 &     0.031 \\
SN1995D  & {\BVRI}  & 12, 13       & 32.71   & 12~ &    0.040 &     0.058 \\
\hline 
\multicolumn{7}{l}{$^a$References:} \\
\multicolumn{7}{l}{
1 - Wells et al.\ \cite{wel94}, 2 - Lira et al.\ \cite{lir98}, 
3 - Filippenko et al.\ \cite{fil92a},} \\ 
\multicolumn{7}{l}{
4 - Leibundgut et al.\ \cite{lei93}, 5 - Turatto et al.\ \cite{tur96a}, 
6 - Suntzeff \cite{sun96},} \\ 
\multicolumn{7}{l}{
7 - Hamuy et al.\ \cite{ham96b}, 8 - Richmond et al.\ \cite{ric95},
9 - Patat et al.\ \cite{pat96},} \\
\multicolumn{7}{l}{
10 - Meikle et al.\ \cite{mei96}, 
11 - Smith et al.\ priv.\ comm., 12 - Riess et al.\ \cite{rie99}, } \\
\multicolumn{7}{l}{
13 - Sadakane et al.\ \cite{sad96}, 14 - Saha et al.\ \cite{sah99}, 
15 - Fisher et al.\ \cite{fis99},} \\
\multicolumn{7}{l}{
16 - Hamuy et al.\ \cite{ham96d} and references therein, 
17 - Suntzeff et al.\ \cite{sun99} and references therein,} \\
\multicolumn{7}{l}{
18 - Riess et al.\ \cite{rie96b} and references therein
}
\end{tabular}
\end{minipage}
\end{table*}
%
%

\section{Fitting filter light curves \label{fitlc}}
\subsection{Method \label{fitmeth}}
The light curves are analyzed using a descriptive model (Vacca \& Leibundgut
\cite{vac96}). For each supernova the observed light curve in each filter is
fit with an empirical model consisting of a Gaussian (for the peak phase) atop
a linear decay (late-time decline), a second Gaussian (to model the secondary
maximum in the {\V}, {\R}, and {\I} band light curves), and an exponentially
rising function (for the pre-maximum segment).  The functional form of the fit
is:

\begin{equation}\label{fiteq}
m = \frac{f_0 + \gamma (t - t_0) 
+ g_0 \exp\left({\frac{-(t-t_0)^2}{2 \sigma_0^2}} \right)
+ g_1 \exp\left({\frac{-(t-t_1)^2}{2 \sigma_1^2}} \right)} 
 {1 - \exp\left({\frac{\tau-t}{\theta}}\right)}
\end{equation}

The first Gaussian and the decline are normalized to the phase $t_0$, while the
second Gaussian occurs at a later phase $t_1$. The exponential cutoff function
for the rise has a characteristic time $\theta$ and a separate phase zero-point
$\tau$. The amplitudes of the two Gaussians, $g_0$ and $g_1$, as well as the
intercept of the line, $f_0$, the slope, $\gamma$, the phases $t_0$ and $t_1$,
and the widths, $\sigma_0$ and $\sigma_1$, the characteristic rise time
$\theta$ and its phase $\tau$ are free parameters in the fit.  Each filter
light curve is fitted individually and independently using a
$\chi^2$-minimization procedure to determine the best-fit values of the
parameters. Other parameters which characterize the shape of the light curve,
such as the time of maximum brightness or $\Delta m_{15}$, can then be derived
from the best-fit model.  Although the model is a completely empirical
description of the general shape of SNe~Ia light curves, we note that
theoretical models (Pinto \& Eastman \cite{pin00a}) predict a Gaussian shape
for the peak in models with constant opacity and \element[][56]{Ni} buried well within
the ejecta.

\begin{figure}[t]
\center{\epsfig{file=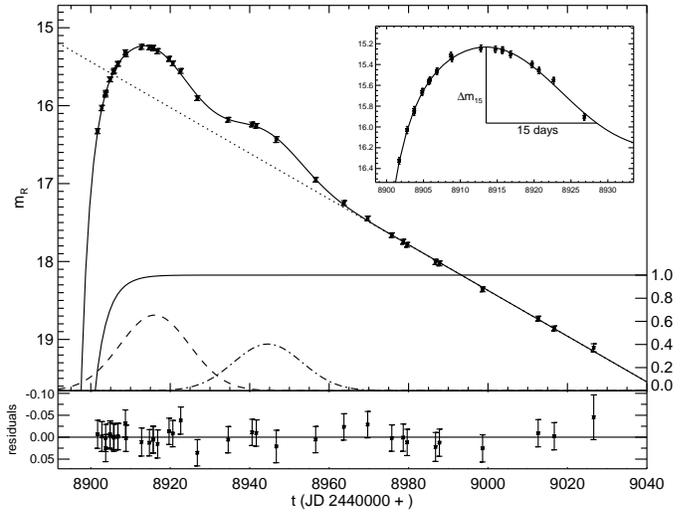,width=\columnwidth}}
\caption{Fit of Eq.~(\ref{fiteq}) to the {\R} photometry of SN~1992bc. The
various components of the fit are displayed as the following: a linear (in
magnitudes) decay (dotted line), an exponential rise factor (solid line) from 0
to 1 as indicated by the right ordinate, two Gaussians for the peak and the
second maximum (dashed and dashed-dotted line) with the amplitude in magnitudes
as indicated by the right ordinate. The inset shows the light curve around
maximum light. {\dm} can be obtained easily because of the continuous
representation of the data by the fit. The residuals from the fit are displayed
in the bottom panel with the observational error bars. \label{fit.ps}}
\end{figure}

Fig.~\ref{fit.ps} shows the example of the fit to the {\R} light curve of
SN~1992bc.
%
%
%
%
The various components of the fit are displayed. The exponential function
(solid line) rises steeply to unity, modeling the rapid brightness increase of
the supernova. The decline rate (dotted line) is set by the long tail beginning
about 50 days past maximum. The first Gaussian (dashed line) fully describes
the maximum phase, as further demonstrated in the inset. The second Gaussian
(dashed-dotted line) reproduces the ``bump'' in the light curve.  The bottom
panel of Fig.~\ref{fit.ps} displays the residuals of the fit with the
observational error bars. It is clear that the function is an accurate,
continuous description of the data.  The small systematic undulations of the
residuals indicate that the fit is not perfect and can be used to make detailed
comparisons between individual supernovae.  We have fitted this model to the
filter light curves of more than 50 supernovae.  These fits will be presented
in forthcoming papers.

The fits produce objective measures of the magnitude and date of maximum, the
extent of the peak phase, the amplitude and extent of the secondary peak, the
late decline rate, and an estimate of the rise time. The accuracy of these
parameters depends strongly on the number and quality of the observations in
each phase. While in most cases the late decline can be derived fairly easily,
the rise time is undetermined when no pre-maximum observations are available.
It is possible to derive other light curve shape descriptions, such as {\dm},
for each filter objectively and independently without comparison to templates.

Other functional forms could be imagined for the fit. However, the late decline
(linear in magnitudes) and the steep rise can be assumed to have simple 
functional
forms. Fitting polynomials or spline functions, for example, could not produce
the transition to the linear decline as observed, but would simply transfer the
undulations of the early phases to the late decline and fail to match the
observed linear decline.  Furthermore, such functions would not provide a small
set of adjustable parameters which can be compared between different objects.

\begin{figure}[t]
\center{\epsfig{file=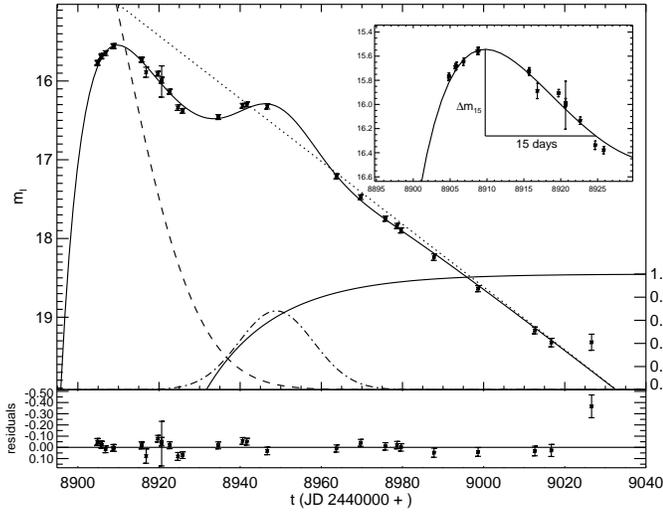,width=\columnwidth}}
\caption{Fit of Eq.~(\ref{fiteq}) to the observed {\I} data of SN~1992bc.
The description of the lines is given in Fig.~\ref{fit.ps}. \label{fitI.ps}}
\end{figure}

Nevertheless, the adopted functional form is not a perfect representation of
all optical filter light curves.  When fitting the {\I} light curves of some
SNe~Ia we found that the model had difficulties to match the observations with
the same accuracy as in the other filters.  This can be seen in
Fig.~\ref{fitI.ps}.
%
%
%
%
The slope of the {\I} band light curve decline (dotted line) is so steep that
the model cannot reproduce the observed light curve.  This steep slope causes
the function to overshoot the observed values near the second maximum and the
fitting program inverts the second Gaussian to produce the observed minimum in
the light curve.  If the fitting routine is constrained to fit only positive
Gaussians, the exponential rising function (solid line) is shifted to late
times and depresses the linear decay (dotted line) before the second maximum;
in addition the first Gaussian (dashed line) is enhanced and shifted to earlier
times.  This implies that, for the {\I} band, the fitted parameters describing
the first Gaussian (i.e.\ its magnitude, its center and its standard deviation)
cannot be used to characterize the first maximum, even if the light curve
appears fitted well.  In spite of this problem, the I band light curves of the
SNe in our sample can all be reasonably well fit by the model and the
data can be represented by the continuous fits. This also means that
the derived light curve parameters, e.g. decline rates, are meaningful.

This steep decline in the {\I} band is a rather unexpected result.  The
physical explanation behind this behavior is that the {\I} filter light curve
is dominated by a rapidly decreasing flux component and significant flux
redistribution takes place in the evolution.  Similar results have been found
by Suntzeff (\cite{sun96}) in the observations of SN~1992A and Pinto \& Eastman
(\cite{pin00b}, see also Eastman \cite{eas97}) in theoretical models.

\subsection{Uncertainties in the model parameters \label{fitunc}}

\begin{figure}[t]
\center{\epsfig{file=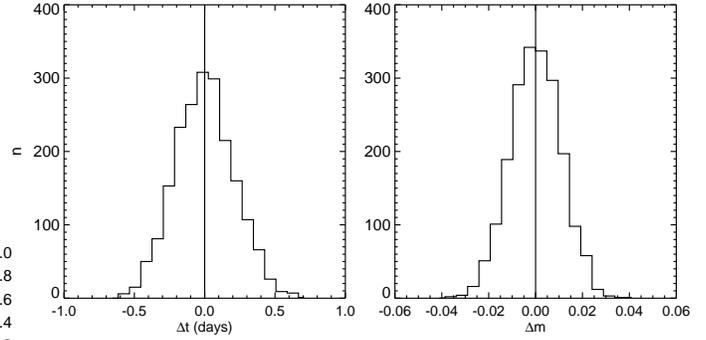,width=\columnwidth}}
\caption{Distributions of the differences between the best fit and the Monte
Carlo simulations for the time of {\B} maximum (left) and {\B} magnitude at
maximum (right) for SN~1992bc.
\label{bmc}}
\end{figure}

The fitting procedure provides an estimate of the goodness-of-fit, as well as
the associated uncertainties on the fit parameters.  Uncertainties on the
derived quantities depend strongly on the quality and number of data points. If
there are no data before 5 days prior to maximum, e.g., the rise time cannot be
determined reliably.  We have estimated the uncertainties using a Monte Carlo
method. We constructed synthetic data sets with the same temporal sampling as
the observed light curve. The magnitude at each point was computed from the
best fit model, with the assumption of a Gaussian probability distribution
whose width was given by the observational uncertainty. The standard deviations
of the observed data points range from 0.03 to 0.14 magnitudes.  In this
manner, 2000 synthetic data sets were simulated. Each synthetic data set was
fit and the frequency distribution for each model parameter was constructed.
Two examples of the resulting distributions (the time of {\B} maximum and the
{\B} magnitude at maximum for SN~1992bc) are shown as histograms in
Fig. \ref{bmc}.  The differences between the mean and the best fit is less than
0.03 days for the time of maximum and less than 0.01 magnitudes for the peak
brightness.  In this example the skewness and kurtosis are negligible. Non-zero
values of skewness and kurtosis provide an indication of the unreliability of
the derived fit parameters.

%
%
%
%
Occasionally the standard deviations obtained for a fit parameter from the
fitting procedure differed significantly from the values obtained from the
Monte Carlo simulations.  This arises because the standard deviations derived
from the fit are calculated from the diagonal elements of the curvature matrix,
while the standard deviations derived from the Monte Carlo simulations result
from actual frequency distributions. (Both standard deviations should be the
same for a perfect light curve, i.e.\ well sampled data, where the brightness
differs from the analytical shape of the light curve by only a small error.
We confirmed our code with this test.)
In all cases we adopted the uncertainties given by the Monte Carlo simulations.

\section{Epochs of maxima \label{epmax}}
\begin{figure}[t]
\center{\epsfig{file=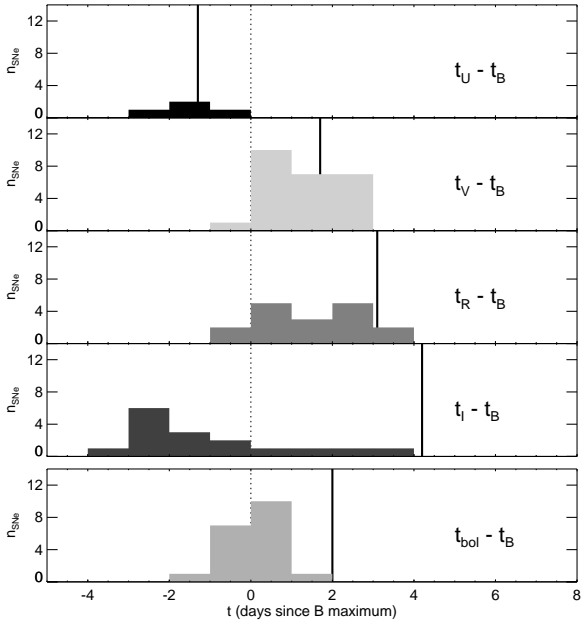,width=\columnwidth}}
\caption[]{Histograms showing the times of the filter maxima from {\U} to {\I}
and of the bolometric luminosity relative to the {\B} maximum. The vertical
lines show the time differences as expected for an expanding, adiabatic cooling
sphere according to Arnett (\cite{arn82}). \label{epochmax} }
\end{figure}

For all 22 SNe~Ia whose light curves were well-sampled around the peak
(SNe~1989B, 1991T, 1991bg, 1992A, 1992bc, 1992bo, 1994D, 1995D, see
Table~\ref{data.tab}; SN~1990N, Lira et al.\ \cite{lir98}; SNe~1990af, 1992P,
1992al, 1992bh, 1992bp, 1992bs, 1993H, 1993O, 1993ag, 
Hamuy et al.\ \cite{ham96b}; SNe~1994S, 1994ae,
1995E, 1995ac, 1995al, 1995bd, 1996X, Riess \cite{rie96c}, Riess et
al. \cite{rie99}) we determined the epochs of individual filter maxima.  A
general trend can be observed in Fig.~\ref{epochmax}, which presents the time
of the filter maxima relative to the epoch of the maximum in {\B}.
%
%
%
%
The {\U} band appears to rise faster than the {\B} band, although the limited
number of objects (4 supernovae) precludes any definitive statements. The {\V}
light curve clearly peaks some time after the {\B} maximum for all observed
SNe~Ia.  This behavior is more or less expected from models of an expanding
cooling atmosphere (Arnett \cite{arn82}), as indicated by the lines in
Fig.~\ref{epochmax}. However, the behavior of the {\R} and {\I} band maxima do
not agree with this simple model; for these bands, the light curve maxima are
reached earlier than in a thermal model. The {\I} histogram is very broad but
for most objects the rise in {\I} is clearly faster than in {\B}. This trend is
further continued in the {\JHK} light curves which all peak before {\B} (Elias
et al.\ \cite{eli85}, Leibundgut \cite{lei88}, Meikle \cite{pme00}).  The early
appearance of the peak in the infrared rules out the idea of an expanding,
cooling sphere.

\section{Bolometric light curves of SNe~Ia \label{constrbol}\label{discussbol}}
The flux emitted by a SN~Ia in the UV, optical, and IR wavelengths, the
so-called ``uvoir bolometric flux'', traces the radiation converted from the
radioactive decays of newly synthesized isotopes. As nearly 80\% of the
bolometric luminosity of a typical SN~Ia is emitted in the range from 3000 to
10000~\AA\ (Suntzeff \cite{sun96}), the integrated flux in the {\UBVRI}
passbands provides a reliable measure of the bolometric luminosity and
therefore represents a physically meaningful quantity. This luminosity depends
directly on the amount of nickel produced in the explosion, the energy
deposition, and the $\gamma$-ray escape, but not on the exact wavelengths of
the emitted photons.

We used the fits of the filter light curves in the {\UBVRI} passbands to
construct an optical bolometric light curve for the SNe in our sample.  All
objects with well-sampled ({\U}){\BVRI} light curves and sufficient coverage
from pre-maximum to late decline phases have been included.  The objects are
listed in Table~\ref{data.tab}. To calculate the absolute bolometric
luminosities, one has to account for reddening and distance moduli; the values
we adopted were taken from the literature as listed in Table \ref{data.tab}. A
galactic extinction law has been employed, as justified by Riess et al.\
(\cite{rie96a}).

\begin{figure}[t]	
\center{\epsfig{file=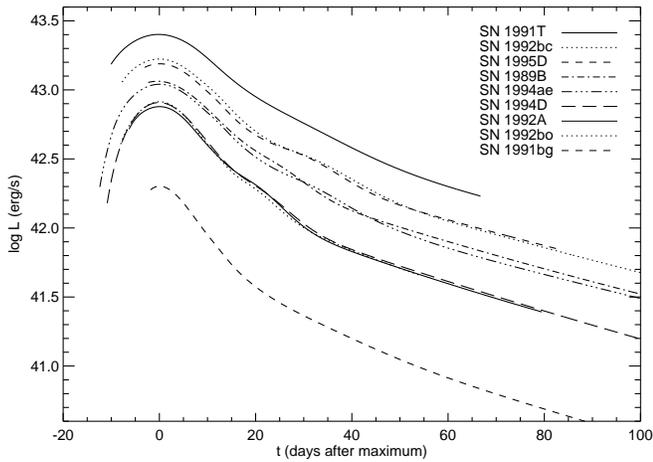,width=\columnwidth}}
\caption[]{Bolometric luminosities for 9 well observed SNe. Only the time range
with the maximum number of filters for each supernova is displayed.  The
reddening and distance moduli have been taken from the literature as listed in
Table \ref{data.tab} and corrections for missing {\U} band are applied as
described in Sect.~\ref{pbcorr}. \label{lumbol}}
\end{figure}

The bolometric light curves of our sample of SNe~Ia are shown in
Fig.~\ref{lumbol}. Only the time range with all available filter photometry is
plotted.  The peak luminosities are clearly different for the objects in the
sample, and these differences are larger than the uncertainties in the
derivation. The most striking feature, however, is the varying strength of the
secondary shoulder, which stems from the {\R} and {\I} light curves (see also
Suntzeff \cite{sun96}).

%
%
%
%

The distribution of the epochs of the bolometric peak luminosity relative 
to the {\B} maximum
is shown at the bottom of Fig.~\ref{epochmax}.  The time of bolometric maximum
suffers from the additional uncertainty that the contributions from the
different wavelength regimes change rapidly before and during the peak phase.
Since we are not including any UV flux in our calculations and SNe~Ia become
optically thin in the UV around this phase, the epoch of the maximum could
actually shift to earlier times than what we measure.  This might explain the
discrepancy with the earlier determination of the bolometric maximum for
SN~1990N and SN~1992A by Leibundgut (\cite{lei96a}) and Suntzeff
(\cite{sun96}), respectively, who included IUE and HST measurements.

\subsection{Uncertainties introduced by the data}
\subsubsection{Missing passbands}
\paragraph{Flux outside {\UBVRI:}}

In our analysis we have neglected any flux outside the optical wavelength
range. In particular, contributions to the bolometric flux from the ultraviolet
(below 3200~\AA) and the infrared above 1~${\mu}$m ({\JHK}) regimes should be
considered in the calculations of the bolometric flux.  Using HST and IUE
spectrophotometry for SN~1990N and SN~1992A, Suntzeff (\cite{sun96}) estimated
the fraction of bolometric luminosity emitted in the UV. He found that the
bolometric light curve is dominated by the optical flux; the flux in the UV
below the optical window drops well below 10\% before maximum.  The {\JHK}
evolution was assumed to be similar to that presented by Elias et al.\
(\cite{eli85}); these passbands add at most 10\% at early times and no more
than about 15\% 80 days after maximum. For example, examination of the data of
SNe~1980N and 1981D (infrared data from Elias et al.\ (\cite{eli81}) and
optical from Hamuy et al.\ (\cite{ham91})) shows that not more than 6\% of the
total flux is emitted beyond {\I} until 50 days past maximum, when the IR data
stop.

\begin{figure}[t]	
\center{\epsfig{file=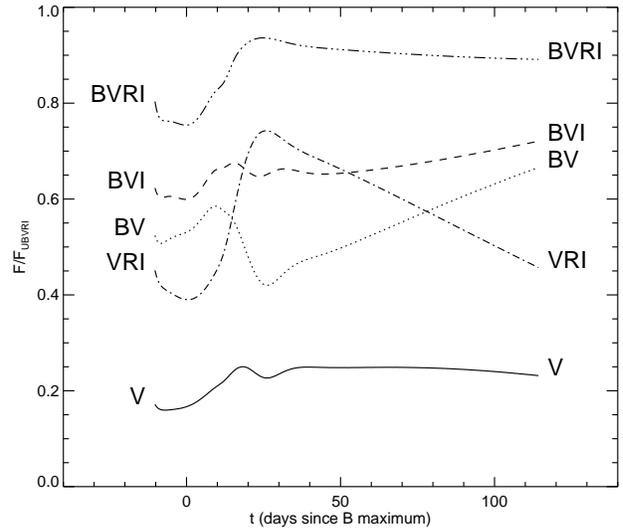,width=\columnwidth}}
\caption[]{Correction factors for missing passbands. The correction factors are
obtained by comparing the flux in the passbands with the total
{\UBVRI}-flux.\label{corr} }
\end{figure}

\paragraph{Corrections for passbands missing in the optical range:
\label{pbcorr}}
For missing passbands between {\U} and {\I} one can infer corrections derived
from those SNe~Ia which have observations in all filters.  Fig.~\ref{corr}
shows the correction factors obtained from SN~1994D. A cautionary note is
appropriate here: SN~1994D displayed some unusual features, in particular a
very blue color at maximum.

%
%
%
%
To estimate the flux corrections we divided the flux in each passband by the
bolometric flux.  Since the filter transmission curves do not continuously
cover the spectrum (i.\ e., there are gaps between {\U} and {\B} as well as
between {\B} and {\V} and overlaps between {\V} and {\R}, and {\R} and {\I})
the coaddition suffers from the interpolation between these passbands.

An interesting result from Fig.~\ref{corr} is the nearly constant bolometric
correction for the {\V} filter. This filter has been used in the past as a
surrogate for bolometric light curves (e.g.\ Cappellaro et al.\ \cite{cap97b}).
We confirm the validity of this assumption for phases between 30 and 110 days
after {\B} maximum where the overall variation is less than 3\%.

\begin{figure}[t]	
\center{\epsfig{file=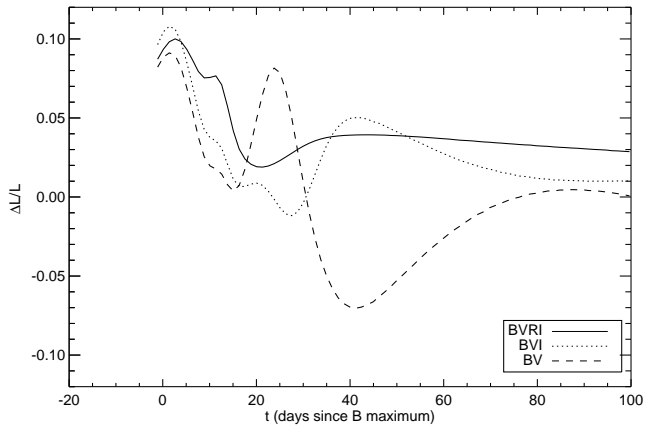,width=\columnwidth}}
\caption[]{Residuals for substituting individual passband observations of
SN~1989B by the corrections derived from SN~1994D.  The errors never exceed
10\% even when more than one filter is missing. \label{testcorr} }
\end{figure}

In order to test our procedure, we calculated bolometric light curves for the
three SNe~Ia which have the full wavelength coverage after purposely omitting
one or more passbands and applying the correction factors we derived from
SN~1994D.  As Fig.~\ref{testcorr} demonstrates, the error is less than 10\% at
all times even if more than one filter is missing, although the errors vary
considerably during the peak and the secondary shoulder phases. The results of
this exercise gave us confidence that we could correct the bolometric light
curves of the remaining six SNe for the missing {\U} band without incurring
large errors.

%
%
%
%

\subsubsection{Effects of systematic differences in photometry sets on
bolometric light curves} Filter light curves from different observatories often
show systematic differences of a few hundredths of a magnitude.  We examined
the effect such systematic errors might have on our bolometric light curves by
artificially shifting individual filter light curves by 0.1 magnitudes and
recomputing the bolometric light curves.  In all cases the effect on the
bolometric light curve is far less than the uncertainties in the distances and
the extinction of the supernovae.  Shifting the {\R} or {\I} light curve by 0.1
magnitude we found that the bolometric luminosity changed by 2\% at maximum and
5\% 25 days later (approximately at second maximum).  For typical systematic
uncertainties of 0.03 magnitudes in all filters, a maximal error of 2 to 3\% is
incurred.

We also constructed the bolometric light curve of SN~1994D for the individual
data sets available (see references to Table~\ref{data.tab}).  The difference
never exceeds 4\% out to 70 days after maximum even though the photometry in
the individual filters differs up to 0.2 magnitudes in {\U}, 0.1 mag in {\BRI}
and 0.05 mag in {\V}.

\subsection{Uncertainties introduced by external parameters}
While uncertainties in the distance moduli will affect the absolute
luminosities in each passband (as well as the bolometric luminosity), the shape
of the light curve is unaffected.  As all distances used here are scaled to a
Hubble constant of $H_0 = 65 \,\mathrm{km\,s^{-1}\,Mpc^{-1}}$, the luminosity
differences are affected only by errors in the determination of the relative
distance modulus.

Reddening, however, affects both the absolute luminosity and the light curve
shape.  The influence of extinction changes as a function of phase with the
changing color of the supernova.  A color excess of {\EBV} $ = 0.05$ decreases
the observed bolometric luminosity at $t = t_{\mathrm{max}}(\mathrm{bol})$ by
15\%, while near the time of the second maximum in the {\R} and {\I} light
curves ($t = 20\,\mathrm{days}$) the observed bolometric luminosity is reduced
by 12\%.  A stronger extinction of {\EBV} $ = 0.35$ reduces the observed
bolometric luminosity by 67\% (56\%) at 
$t = t_{\mathrm{max}}$ ($t =20\,\mathrm{days}$).

The uncertainty in the reddening estimate introduces subtle additional
effects. If $\delta${\EBV}$= 0.02$~mag at low reddenings (\EBV $< 0.05$), an
additional uncertainty of 5\% in the bolometric luminosity is introduced.  At
higher reddening values ({\EBV} $ = 0.3$), uncertainties of $\delta${\EBV}=0.05
and 0.10 produce
changes of 15\% and 31\% in the bolometric luminosity, respectively.

The decline rate of the bolometric light curve decreases with increasing
reddening due to the color evolution of the supernova and the selective
absorption.
For SN~1994D {\dm}(Bol) would evolve linearly from 1.13 at hypothetical {\EBV}
$ =0$ to 0.99 at {\EBV} $ =0.35$.  The linearity breaks down at about {\EBV} $
\approx 0.5$.

As described by Leibundgut (\cite{lei88}) the reddening depends on the
intrinsic color of the observed object (Schmidt-Kaler \cite{sch82}). This
implies that the color evolution influences the shape of the filter light
curves.  A color difference of {\BmV}$ = 0.7$ for SNe~Ia in the first 15 days
results in an increase of {\dm}({\B}) by $0.2 \times {\EBV}$ simply due to the
color dependence of the reddening.  The increase of {\dm} for blue filters is
larger than for the redder passbands.

\subsection{Uncertainties introduced by the method}
\begin{figure}[t]	
\center{\epsfig{file=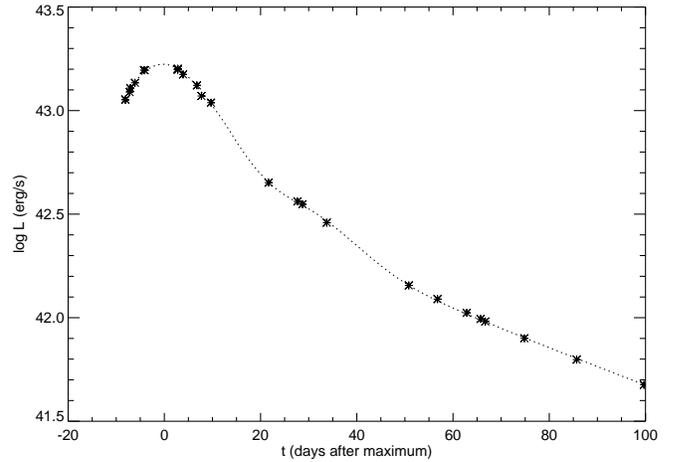,width=\columnwidth}}
\caption[]{ 
Constructing the bolometric light curves before and after fitting the filter
light curves of SN~1992bc. The stars indicate the bolometric luminosity as
calculated directly from the observations. The correction for missing {\U} band
is applied as described in Sect.~\ref{pbcorr}, only the epochs with the maximum
number of filters are shown. The dotted line is the bolometric luminosity
calculated from the fits to the individual passbands.
\label{bolfldat} }
\end{figure}

The effect of fitting the light curves before constructing the bolometric light
curves can be seen in Fig.~\ref{bolfldat}. The bolometric flux for SN~1992bc
determined in this way is compared against the straight integration over the
wavelengths of the filter observations. The agreement between the two
approaches is excellent and no differences can be observed. This also applies
to the correction for the missing {\U} filter observations.  Deviations can be
found only for those parameters derived from the light curve, which extrapolate
far from the observed epochs, e.g.\ the rise time.
%
%
%
%

The integration over wavelengths to calculate the bolometric light curve can be
performed in different ways. A straight integration of the broad-band fluxes
has to take into account the transmission of and any overlaps and gaps between 
filters. We experimented with several integration techniques, but found that
for all interpolation methods, we reproduced the bolometric flux to within 2\%
at any epoch we considered. This has already been shown for SN~1987A by
Suntzeff \& Bouchet (\cite{sun90}). We chose to calculate the bolometric 
flux by summing the flux at the effective filter wavelength multiplied with
the filter bandwidth. 

\section{Discussion\label{disc}}

\subsection{Peak Bolometric Flux}

The bolometric light curves presented in Fig.~\ref{lumbol} were constructed
from {\UBVRI} light curves where available (SN~1989B, SN~1991T and SN~1994D).
All others are based on {\BVRI} light curves with a correction for the missing
{\U} band applied as described in Sect.~\ref{pbcorr}. The luminosities depend
directly on the assumed distance moduli to the individual supernovae. We have
used the current best estimates from the literature (Table \ref{data.tab}), but
some of the distance estimates may change when more accurate distances become
available.  We also corrected the magnitudes for extinction, which, for some
objects, can be fairly substantial (SN~1989B, SN~1991T, and SN~1994ae). From
Table~\ref{tab:lum} it is clear that our sample displays a rather large range
in luminosities, both in {\B} and the bolometric maximum. SN~1991bg is 11 times
less luminous in {\B} than the brightest SN~Ia in the sample, SN~1991T. For the
bolometric maximum we find a factor of 12.3.  Excluding this peculiar object we
still find a range in luminosity of a factor 4.7 in the filter passband and a
factor of 3.3 in the bolometric peak flux.

\subsection{Light Curve Shape}

The shapes of the bolometric light curves are unaffected by the distance
modulus and are only marginally influenced by reddening (see
Sect.~\ref{discussbol}).  The most striking feature in these light curves is
the inflection near 25 days past maximum (Suntzeff \cite{sun96}).  It is
observed in all SNe~Ia, with the notable exception of SN~1991bg
(Fig.~\ref{lumbol}).  The strength of this shoulder varies from a rather weak
flattening in the bolometric light curve of SN~1991T to a very strong bump in
SN~1994D. This bump arises from the strong secondary maximum in the {\R} and
{\I} light curves. The secondary maximum is also observed in the near-IR {\JHK}
light curves (Elias et al.  \cite{eli85}, Meikle \cite{pme00}, Meikle \&
Hernandez \cite{mei00}) which have not been used for the calculation of
bolometric light curves presented here.

%
%
\begin{table*}
\caption{Absolute {\B} magnitudes and bolometric luminosities. The
bolometric luminosities have been corrected for the missing {\U} band
where appropriate. The nickel mass is derived from the luminosity for a
rise time of 17 days to the bolometric peak.
\label{tab:lum}} 
\begin{tabular}{l l r r r r r r} 
SN & ~~$M_{\B}$ & $\Delta m_{15}^{\B}$ & $\log L_\mathrm{bol}$ & $\Delta
m_{15}^\mathrm{bol}$ & 
$M_{\mathrm{Ni}}$~ & $t_{-1/2}$ & $t_{+1/2}$ \\
  &   (mag)  & (mag)  &   (erg s$^{-1}$) & (mag)  &  ($M_{\sun}$) & (days) &
  (days) \\ 
(1) & ~~~(2) & (3)~ & (4)~~~ & (5)~~ & (6)~~ & (7)~~ & (8)~~ \\
\hline 
SN1989B  & -19.37 & 1.20~ & 43.06~~ & 0.91~ &   0.57~ & ~---~~ & 13.1~ \\
SN1991T  & -20.06 & 0.97~ & 43.36~~ & 0.83~ &   1.14~ & ~11.6~ & 14.0~ \\
SN1991bg & -16.78 & 1.85~ & 42.32~~ & 1.42~ &   0.11~ & ~---~~ & ~8.8~ \\
SN1992A  & -18.80 & 1.33~ & 42.88~~ & 1.15~ &   0.39~ & ~~8.6~ & 10.3~ \\
SN1992bc & -19.72 & 0.87~ & 43.22~~ & 0.93~ &   0.84~ & ~10.1~ & 13.0~ \\
SN1992bo & -18.89 & 1.73~ & 42.91~~ & 1.27~ &   0.41~ & ~~8.3~ & ~9.6~ \\
SN1994D  & -18.91 & 1.46~ & 42.91~~ & 1.16~ &   0.41~ & ~~7.5~ & 10.8~ \\
SN1994ae & -19.24 & 0.95~ & 43.04~~ & 0.97~ &   0.55~ & ~~9.6~ & 12.6~ \\
SN1995D  & -19.66 & 0.98~ & 43.19~~ & 1.00~ &   0.77~ & ~---~~ & 12.2~ \\
\hline 
\end{tabular}
\end{table*}
%
%

Inspection of Fig.~\ref{lumbol} clearly shows the brighter SNe having wider
primary peaks than the fainter SNe.

In order to quantify this statement we have measured the width of the
bolometric light curve at half the peak luminosity (with $t_{-1/2}$ denoting
the time it takes to rise and $t_{+1/2}$ to decline, see Table \ref{tab:lum}).
There are sufficient pre-maximum observations available for
six supernovae (SN~1991T, SN~1992A, SN~1992bc, SN~1992bo, SN~1994D, and
SN~1994ae) from which the full width of the peaks can be reliably determined.
In three cases the first observations were obtained well below the half-maximum
flux level; in the case of SN~1991T we extrapolated our bolometric light curve
by about 1.5 days and for SN~1992A and SN~1992bc by 2.5 days. The bright
SN~1991T and SN~1992bc show a peak width of 26 and 23 days, respectively, the
intermediate SN~1994ae one of 22 days, while the fainter SN~1994D,
SN~1992A, and SN~1992bo remained brighter than half their maximum luminosity
for about 18 to 19 days.

The pre-maximum rise $t_{-1/2}$ in the bolometric light curve is, in all cases,
substantially faster than the decline $t_{+1/2}$ by between 20 to 30\% or a
time difference of between 2 to almost 4 days (Table~\ref{tab:lum}).

\begin{figure}[t]	
\center{\epsfig{file=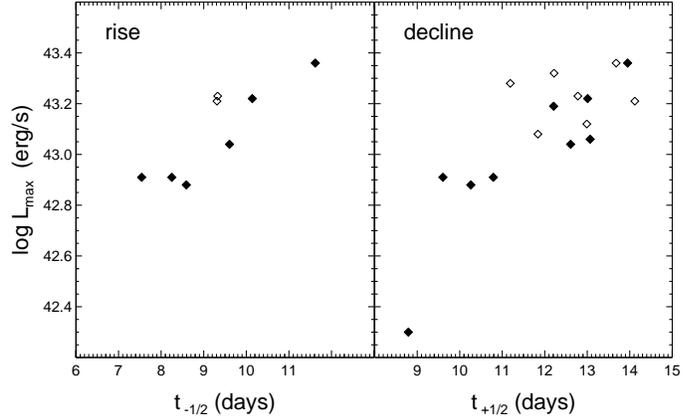,width=\columnwidth}}
\caption[]{Rise to and decline from maximum in bolometric light curves.  The
time between maximum luminosity and half its value is plotted. The sample of
SNe was extended to the one used in Sect.~\ref{epmax}, the SNe from
Table~\ref{data.tab} are shown as filled symbols. Only SNe with sufficient
coverage (rise: first observation not later than 3 days after $t_{-1/2}$,
decline: first observation before $t_{\mathrm{max}}$(Bol)) have been included.
\label{thalf} }
\end{figure}

The decline to half the supernova's peak luminosity $t_{+1/2}$ varies from 9
days to 14 days.  A weak correlation between maximum brightness and decline
rate can be seen (Fig.~\ref{thalf}).  The sample of SNe was extended to the one
used in Sect.~\ref{epmax}. In Fig.~\ref{thalf} only those SNe with sufficient
time coverage are shown.

%
%
%
%

Significant coverage of the rise to maximum is available for only two
supernovae.  There are too few SNe to make a definitive statement here about
the rise times of SNe~Ia in general.  Nevertheless, our formal fits to SN~1994D
and SN~1994ae give rise times of 16.4 and 18.2 days, respectively, for their
bolometric light curves. The formal errors in this parameter, derived with the
Monte Carlo algorithm described in Sect.~\ref{fitunc}, is 2 days for SN~1994D
and 1 day for SN~1994ae.  If the bolometric luminosities are calculated by
fitting the bolometric light curve from the individual observations, the
derived rise times are slightly different ($18$ and $19$ days, respectively).
A previous analysis of SN~1994D derived a rise time for the bolometric light
curve of 18 days (Vacca \& Leibundgut \cite{vac96}) in fair agreement with our
current analysis.

These rise times are significantly larger than most light curve calculation of
explosion models (H\"oflich et al.\ \cite{hoe96}, Pinto \& Eastman
\cite{pin00a}).  These theoretical calculations typically yield rise times of
15 days.

The secondary bump clearly indicates a change in the emission mechanism of
SNe~Ia. Its varying strength might be related to the release of photons
``stored'' in the ejecta (the ``old'' photons described in Pinto \& Eastman
\cite{pin00b} and Eastman \cite{eas97}).  In such a scenario the bump arises
from the fact that the ejecta become optically thin during this phase, due to
subtle differences in the ejecta structure and ejecta velocity.

The exponential decline of the bolometric light curves between 50 and 80 days
past bolometric maximum is remarkably similar for all supernovae. The decline
at this phase is still dominated by the \element[][56]{Co} decay and the
$\gamma-$ray escape fraction (Milne et al.  \cite{mil99}).  We find a decline
rate of 2.6$\pm$0.2 mag per 100~days. The sample is still very small but
SN~1991bg stands out with a decline rate of 3.0~mag/(100~days). The uniformity
of the late declines indicates that the differences observed at earlier phases
are due to photospheric effects when the optical depth for optical radiation is
large and not because of the explosion mechanism. In particular the fraction of
energy converted to optical radiation at late phases appears to change in an
identical fashion for the majority of the objects, although SN~1991bg proves to
be the exception to the rule once again.  The change in the column density in
the different supernovae must be very similar, despite the different
luminosities at these epochs. This indicates that the kinetic energy is somehow
coupled to the pre-supernova mass.

The decline rate during the same period in the {\V} filter is identical
with 2.6$\pm$0.2 mag/(100~days) as is also confirmed by the constant fraction
of the {\V} to the bolometric flux (cf.  Fig.~\ref{corr}).  These decline rates
have been shown to correlate very well with measurements derived by others
(Vacca \& Leibundgut \cite{vac97}).  The comparison with other determinations
of the decline rate is difficult as the slope of the decline continues to vary
even at late epochs (e.g. Suntzeff \cite{sun96}, Turatto et al.\
\cite{tur96a}). This dependence on the epochs observed has to be considered
when a comparison is attempted. In most cases the decline was estimated out to
phases of 200 days (Turatto et al.\ \cite{tur90}), which leads to smaller
decline rates than we find here. In the case of SN~1991bg the decline rate was
measured at early phases, which led to a significantly steeper decline estimate
(Filippenko et al.\ \cite{fil92a}).

Note that the phase range of bolometric light curves presented in this
paper is well before any effects
of positron escape could be measured (Milne et al.\ \cite{mil99}) or the IR
catastrophe takes place (Fransson et al.\ \cite{fra96}).

\subsection{Nickel mass}
Given the peak luminosity it is straight forward to derive the nickel mass
which powers the supernova emission. Near maximum light the photon escape
equals the instantaneous energy input and is directly related to the total
amount of \element[][56]{Ni} synthesized in the explosion (Arnett \cite{arn82},
Arnett et al.\ \cite{arn85}, Pinto \& Eastman \cite{pin00a}). We have used our
bolometric peak luminosities to derive the nickel masses (Table~\ref{tab:lum}).
Since we are not sampling all the emerging energy from the SNe~Ia, but are
restricted to the optical fluxes, we are underestimating the total luminosity.
Suntzeff (\cite{sun96}) estimated that about 10\% are not accounted by the
optical filters near maximum light.  All masses thus have to be increased by a
factor 1.1. Another uncertainty is the exact rise time, which is an important
parameter in the calculation. We have assumed a rise time of 17 days to the
bolometric maximum for all supernovae.  It is likely that there are significant
differences in the rise times and this would alter the estimates for the Ni
mass.  A longer rise time would imply a larger nickel mass for a given measured
luminosity. Decreasing the rise time to 12 days yields only 70\%
$M_{\mathrm{Ni}}$ of the values given in Table~\ref{tab:lum}. Such a short rise
time is excluded for most of the SNe~Ia, where observations as early as 14 days
have been recorded (SN~1990N: Leibundgut et al.\ \cite{lei91c}, Lira et al.\ 
\cite{lir98}, SN~1994D: Vacca \& Leibundgut \cite{vac96}), but could still be
feasible for SN~1991bg. For a more realistic range of 16 to 20 days between
explosion and bolometric maximum the nickel mass would change by only $\mp$10\%
from the values provided here. Clearly, the dominant uncertainty in the
determination of the nickel mass stems from the uncertainties in the distances
and the extinction corrections.

For a few of the supernovae, nickel masses have been measured by other methods.
SN~1991T has an upper limit for the radioactive nickel produced in the
explosion of about 1~$M_{\sun}$ based on the 1.644$\mu$m Fe lines (Spyromilio
et al.\ \cite{spy92}).  This value depends on the exact ionization structure of
the supernova one year after explosion and a conservative range of 0.4 to 1
$M_{\sun}$ had been derived. This is fully consistent with our estimate.  
Bowers et
al.\ (\cite{bow97}) have derived nickel masses for several SNe~Ia in a similar
way. Their best estimates for SN~1991T, SN~1994ae, and SN~1995D are all about
half the value found here, when converted to their distances and extinctions.
However, they point out that their values should be increased by a factor of
1.2 to 1.7 to account for ionization states not included in their analysis.
With this correction we find a good agreement.  Cappellaro et al.\ 
(\cite{cap97b}) derived masses from the late {\V} light curves.  There are four
objects in common with our study: SN~1991bg, SN~1991T, SN~1992A, and SN~1994D.
Adjusting the determinations to the same distances and re-normalizing to our
SN~1991T Ni mass, we find a general agreement, although there are differences
at the 0.1 $M_{\sun}$ level.

Nickel masses were also derived from the line profiles of [Fe~II] and [Fe~III]
lines in the optical by Mazzali et al.\ (\cite{ma98}). These measurements
depend critically on the ionization structure in the ejecta and had been
normalized to a nickel mass of SN~1991T of 1~$M_{\sun}$. When we scale their
masses to our measurement we find a reasonable agreement.

The nickel masses derived in our analysis are well within the bounds of the
current models for SN~Ia explosions (H\"oflich et al.\ \cite{hoe96}, Woosley \&
Weaver \cite{woo94}). There seems to be no real difference in \element[][56]{Ni}
produced by the various explosion models and hence a distinction by the
bolometric light curve alone is not possible.

\section{Conclusions\label{conc}}
 
Fitting light curves with a descriptive model has a number of advantages over
template methods. It is suited to explore the variety among SN~Ia and
provides an independent way to look for correlations.  A simple application of
our fitting procedure has been presented to demonstrate the complicated nature
of SN~Ia emission through the occurrence of the peak luminosity in individual
filters.  The most obvious signatures for non-thermal emission from SNe~Ia have
so far been the infrared light curves (Elias et al.  \cite{eli85}) and the lack
of emission near 1.2$\mu$m (Spyromilio et al. \cite{spy94}, Wheeler et al.\
\cite{whe98}).  It had also been demonstrated through spectral synthesis
calculations (H\"oflich et al.\ \cite{hoe96}, Eastman \cite{eas97}).  Although
there is a fairly large scatter of about 2 days in the relative epoch of filter
maximum light a clear trend to earlier maxima in {\R} and {\I} is observed. In
fact, the {\I} maximum occurs clearly {\it before} the {\B} maximum, a trend
also observed in the {\JHK} light curves (Elias et al.\ \cite{eli85},
Meikle \cite{pme00}).

Another application of the continuous approximation of the observations is the
construction of bolometric light curves.  Bolometric light curves form an
important link between the explosion models and the radiation transport
calculations for SNe~Ia ejecta. We have demonstrated that the effects of
missing passbands, distance modulus, reddening, light curve fitting and the
integration methods are not critical to the shape.  Amongst these, the
uncertainties from the distance modulus towards individual supernovae dominate.

The shapes of the bolometric light curve of individual SNe~Ia vary
significantly. The secondary maxima observed in the {\R} and {\I} light curves
show up with varying strength in the bolometric light curves as well. The
variety of light curve shapes indicates subtle variations in the energy release
of these explosions.  Pinto \& Eastman (\cite{pin00b}) try to explain these
secondary maxima as due to the distribution of material in the inner core of
the explosion, possibly connected to the explosion mechanism. If this is the
case, then the detailed study of the bolometric light curves and their
differences among individual SNe~Ia might provide a direct ``window'' into the
explosion.

With the currently best available distances we find that the peak luminosities
of the SNe~Ia in our sample display a rather large range.  They imply a factor
of 2.5 in the \element[][56]{Ni} masses.  SN~1991bg produced about 8 times less
\element[][56]{Ni} than the brightest object, SN~1992bc. The range of nickel masses
indicates significant differences in the explosions of SNe~Ia.  From the
late-phase decline, we find that the change of the decline rate is rather
uniform indicating similarity in change of the $\gamma-$ray escape fraction for
all SNe~Ia independent of the amount of nickel produced in the explosion.

\begin{acknowledgements}
We are grateful to Leon Lucy for help with the Monte Carlo analysis. We
are indebted to Adam Riess for pointing out an inconsistency in the
adopted distances of an earlier draft.
\end{acknowledgements}


\begin{thebibliography}{}

\bibitem[{1982}]{arn82}
{Arnett} W.~D., 1982, ApJ 253, 785

\bibitem[{1985}]{arn85}
{Arnett} W.~D., {Branch} D.,  {Wheeler} J.~C., 1985, Nat 314, 337

\bibitem[{1997}]{bow97} 
Bowers E.~J.~C., Meikle W.~P.~S., Geballe T.~R., et~al., 1997, MNRAS 290, 663 

\bibitem[{1992}]{bra92}
{Branch} D.,  {Tammann} G.~A., 1992, ARA\&A 30, 359

\bibitem[{1997}]{cap97b}
{Cappellaro} E., {Mazzali} P.~A., {Benetti} S., et~al., 1997, A\&A 328, 203

\bibitem[{1985}]{dog85}
Doggett J.~B., Branch D., 1985, AJ 90, 2303

\bibitem[{1997}]{eas97}
Eastman R.~G., 1997, In: Thermonuclear Supernovae, Ruiz-Lapuente P., 
Canal R., Isern J.\ (eds.), Kluwer, Dordrecht, p.~571

\bibitem[{1981}]{eli81}
Elias J.~H., Frogel J.~A., Hackwell J.~A., et~al., 1981, ApJ 251, L13

\bibitem[{1985}]{eli85}
Elias J.~H., Matthews K., Neugebauer G., et~al., 1985, ApJ 296, 379

\bibitem[{1992}]{fil92a}
{Filippenko} A.~V., Richmond M.~W., Branch D., et~al., 1992, AJ 104, 1543

\bibitem[{1999}]{fis99}
{Fisher} A., {Branch} D., {Hatano} K., et~al., 1999, MNRAS 304, 67

\bibitem[{1996}]{fra96}
Fransson C., Houck J., Kozma C., 1996, In:  IAU
Colloquium 145: Supernovae and Supernova Remnants, McCray R., Wang Z.\ (eds.),
Cambridge University Press, Cambridge, p.~211

\bibitem[{1987}]{fro87}
{Frogel} J.~A., {Gregory} B., {Kawara} K., et~al., 1987, ApJ 315, L129

\bibitem[{1998}]{gar98}
{Garnavich} P.~M., Kirshner R.~P., Challis P., et~al., 1998, ApJ 493, L53

\bibitem[{1991}]{ham91}
Hamuy M., Phillips M.~M., Maza J., et~al., 1991, AJ 102, 208

\bibitem[{1996a}]{ham96d}
{Hamuy} M., {Phillips} M.~M., Schommer R.~A., et~al., 1996a, AJ 112, 2391

\bibitem[{1996b}]{ham96b}
{Hamuy} M., Phillips M.~M., Suntzeff N.~B., et~al., 1996b, AJ 112, 2408

\bibitem[{1996}]{hoe96}
{H\"oflich} P., {Khokhlov} A., {Wheeler} J.~C., et~al., 1996, ApJ 472, L81

\bibitem[{1997}]{hoe97} 
H\"oflich P., Khokhlov A., Wheeler J., et~al., 1997, In: Thermonuclear
Supernovae, Ruiz-Lapuente P., Canal R., Isern J.\ (eds.), Kluwer, Dordrecht,
p.~659 

\bibitem[{1993}]{kho93}
{Khokhlov} A., {M\"uller} E., {H\"oflich} P., 1993, A\&A 270, 223

\bibitem[{1988}]{lei88}
Leibundgut B., 1988, Ph.D.\ thesis, Universit\"at Basel

\bibitem[{1996}]{lei96a} 
Leibundgut B., 1996, In: IAU Colloquium 145: Supernovae and Supernova Remnants,
McCray R., Wang Z.\ (eds.), Cambridge University Press, Cambridge, p.~11

\bibitem[{1992}]{lei92}
{Leibundgut} B., {Pinto} P.~A., 1992, ApJ 401, 49

\bibitem[{1991a}]{lei91c}
{Leibundgut} B., {Kirshner} R.~P., {Filippenko} A.~V., et~al., 1991a, ApJ 371,
L23 

\bibitem[{1991b}]{lei91b}
{Leibundgut} B., {Tammann} G.~A., {Cadonau} R., et~al., 1991b, A\&AS 89, 537

\bibitem[{1993}]{lei93}
{Leibundgut} B., Kirshner R.~P., Phillips M.~M., et~al., 1993, AJ 105, 301

\bibitem[{1998}]{lir98}
{Lira} P., Suntzeff N.~B., Phillips M.~M., et~al., 1998, AJ 115, 234

\bibitem[{1998}]{ma98}
Mazzali P.~A., Cappellaro E., Danziger I.~J., et~al., 1998, ApJ 499, L49

\bibitem[{2000}]{pme00}
{Meikle} W.~P.~S., 2000, MNRAS 314, 782

\bibitem[{2000}]{mei00}
Meikle W.~P.~S., Hernandez M., 2000, In: Future Directions
of Supernova Research, Cassisi S.\, Mazzali P.\ (eds.), Mem.\ Soc.\
Astron.\ Ital., in press (astro-ph/9902056)

\bibitem[{1996}]{mei96}
{Meikle} W.~P.~S., Cumming R.~J., Geballe T.~R., et~al., 1996, MNRAS 281, 263 

\bibitem[{1999}]{mil99}
Milne P.~A., The L.-S., Leising M., 1999, ApJS 124, 503

\bibitem[{1964}]{min64}
Minkowski R., 1964, ARA\&A 2, 247

\bibitem[{1996}]{pat96}
{Patat} F., {Benetti} S., {Cappellaro} E., et al., 1996, MNRAS 278, 111

\bibitem[{1997}]{per97}
{Perlmutter} S., Gabi S., Goldhaber G., et~al., 1997, ApJ 483, 565

\bibitem[{1993}]{phi93}
{Phillips} M.~M., 1993, ApJ 413, L105

\bibitem[{1987}]{phi87}
{Phillips} M.~M., Phillips A.~C., Heathcote S.~R., et~al., 1987, PASP 99, 592

\bibitem[{1999}]{phi99}
Phillips M.~M., Lira P., Suntzeff N.~B., et~al., 1999, AJ 118, 1766

\bibitem[{2000a}]{pin00a}
Pinto P.~A., Eastman R.~G., 2000a, ApJ 530, 744

\bibitem[{2000b}]{pin00b}
Pinto P.~A., Eastman R.~G., 2000b, ApJ 530, 757

\bibitem[{1995}]{ric95}
{Richmond} M.~W., Treffers R.~R., Filippenko A.~V., et~al., 1995, AJ 109, 2121 

\bibitem[{1996}]{rie96c}
Riess A.~G., 1996, Ph.D.~thesis, Harvard University, Cambridge, MA

\bibitem[{1996a}]{rie96b}
{Riess} A.~G., {Press} W.~H., {Kirshner} R.~P., 1996a, ApJ 473, 88

\bibitem[{1996b}]{rie96a}
{Riess} A.~G., {Press} W.~H., {Kirshner} R.~P., 1996b, ApJ 473, 588

\bibitem[{1998a}]{rie98a}
Riess A.~G., Filippenko A.~V., Challis P., et~al., 1998a, AJ 116, 1009

\bibitem[{1998b}]{rie98b}
Riess A.~G., Nugent P., Filippenko A.~V., et~al., 1998b, ApJ 504, 935

\bibitem[{1999}]{rie99}
Riess A.~G., Kirshner R.~P., Schmidt B.~P., et~al., 1999, AJ 117, 707

\bibitem[{1996}]{sad96}
{Sadakane} K., Yokoo T., Arimoto J.-I., et~al., 1996, PASJ 48, 51

\bibitem[{1999}]{sah99}
Saha A., Sandage A., Tammann G.~A., et~al., 1999, ApJ 522, 802

\bibitem[{1995}]{schle95}
Schlegel E.~M., 1995, AJ 109, 2620

\bibitem[{1998}]{schle98}
Schlegel D.~J., Finkbeiner D.~P., Davis M., 1998, ApJ 500, 525

\bibitem[{1982}]{sch82}
Schmidt-Kaler T., 1982, In: Landolt-B\"ornstein
Vol.~VI/2b, Schaifer K., Voigt H.\ (eds.), Springer, Berlin, p.~10

\bibitem[{1998}]{sch98}
Schmidt B.~P., Suntzeff N.~B., Phillips M.~M., et~al., 1998, ApJ 507, 46

\bibitem[{1992}]{spy92}
Spyromilio J., Meikle W.~S.~P., Allen D.~A., et~al., 1992, MNRAS 258, 53

\bibitem[{1994}]{spy94}
Spyromilio J., Pinto P.~A., Eastman R.~G., 1994, MNRAS 266, L17

\bibitem[{1996}]{sun96}
Suntzeff N.~B., 1996, In: IAU Colloquium 145: Supernovae and Supernova
Remnants, McCray R., Wang Z.\ (eds.), Cambridge University Press,
Cambridge, p.~41

\bibitem[{1990}]{sun90}
{Suntzeff} N.~B., {Bouchet} P., 1990, AJ 99, 650

\bibitem[{1999}]{sun99}
Suntzeff N.~B., Phillips M.~M., Covarrubias R., et~al., 1999, AJ 117, 1175

\bibitem[{1990}]{tur90}
Turatto M., Cappellaro E., Barbon R., et~al., 1990, AJ 100, 771

\bibitem[{1996}]{tur96a}
{Turatto} M., {Benetti} S., {Cappellaro} E., et~al., 1996, MNRAS 283, 1

\bibitem[{1996}]{vac96}
{Vacca} W.~D., {Leibundgut} B., 1996, ApJ 471, L37

\bibitem[{1997}]{vac97} 
Vacca W.~D., Leibundgut B., 1997, In: Thermonuclear Supernovae, Ruiz-Lapuente
P., Canal R., Isern J.\ (eds.), Kluwer, Dordrecht, p.~65

\bibitem[{1994}]{wel94}
{Wells} L.~A., Phillips M.~M., Suntzeff N.~B., et~al., 1994, AJ 108, 2233

\bibitem[{1998}]{whe98}
Wheeler J.~C., H\"oflich P., Harkness R.~P., et~al., 1998, ApJ 496, 908

\bibitem[{1994}]{woo94}
Woosley S.~E., Weaver T.~A., 1994, ApJ 423, 371

\end{thebibliography}
\end{document}